\documentclass[reprint,prl,superscriptaddress,nofootinbib]{revtex4-1}

\usepackage{graphicx}
\usepackage{dcolumn}
\usepackage{bm}
\usepackage{float}
\usepackage{amsthm,amssymb,amsmath,url,braket,amsbsy} 
\usepackage{threeparttable}
\usepackage[utf8]{inputenc}
\usepackage{dirtytalk}
\graphicspath {{figures/}}
\usepackage{color}
\usepackage{cancel}

\begin{document}

\title{Magnetization direction dependent spin Hall effect in 3\textit{d} ferromagnets}

\author{Guanxiong Qu}
\affiliation{Department of Physics Engineering, The University of Tokyo, Tokyo 113-0033, Japan}

\author{Kohji Nakamura}

\affiliation{Department of Physics Engineering, Mie University, Tsu, Mie 514-8507, Japan}

\author{Masamitsu Hayashi}

\affiliation{Department of Physics Engineering, The University of Tokyo, Tokyo 113-0033, Japan}

\date{\today}

\begin{abstract}
We have studied the intrinsic spin Hall conductivity in 3\textit{d} transition metal ferromagnets using first principle calculations.
We find the spin Hall conductivity of bcc-Fe and fcc-Ni, prototypes of ferromagnetic systems, depends on the direction of magnetization. 
The spin Hall conductivity of electrons with their spin orientation orthogonal to the magnetization are found to be larger than that when the two are parallel. 
For example, the former can be more than four times larger than the latter in bcc-Fe. 
Such difference arises due to the anisotropy of the spin current operator in the spinor space: its expectation value with the Bloch states depends on the relative angle between the conduction electron spin and the magnetization.
A simple analytical form is developed to describe the relation between the spinor states and the Berry curvature and the spin Berry curvature.
The form can account for the features found in the calculations. 
These results show that ferromagnets can be used to generate spin current and its magnitude can be controlled by the magnetization direction. 
\end{abstract}


\maketitle

\section{Introduction}
The spin Hall effect (SHE)\cite{Sinova2015} allows generation of spin current when current is passed to materials with large spin orbit coupling (SOC). Such spin current can be used to manipulate the magnetization direction of a nearby ferromagnet using the spin transfer torque, or the so-called spin orbit torque\cite{Liu2012}. Giant spin Hall effect has been found in non-magnetic 5\textit{d} transition metals\cite{Liu2012a}. Search for materials with large spin Hall effect has been expanded into various systems, including ferromagnets\cite{Miao2013,Tsukahara2014} and antiferromagnets\cite{Fukami2016,Zhang2016}. 

Model calculations\cite{Taniguchi2015} have predicted that ferromagnets can be used to generate spin accumulation through its anomalous Hall effect (AHE). Experimentally it has been shown that the spin accumulation from the AHE in ferromagnets can be used to manipulate the magnetic moments of a nearby ferromagnetic or ferrimagnetic layer\cite{Tian2016,Das2017,Iihama2018,Baek2018,kimata2019}. Under such circumstance, the degree of spin accumulation from the AHE can be tuned by the magnetization direction of the ferromagnetic layer. The spin Hall effect in ferromagnets, in contrast, is not well understood partly because of the difficulty in distinguishing spin current from the AHE and SHE. In theory, it has been reported that in ferromagnets, spin current with spin polarization transverse to the magnetization is protected from dephasing\cite{amin2019}. Recent experiments, however, indicate contradictory pictures on the SHE in ferromagnets, suggesting that it can either be dependent or independent of the magnetization direction\cite{Tian2016,Das2017}. Clarifying the underlying physics of SHE in ferromagnets thus remains as a challenge.

With the emergence of topology in condensed matter physics, the AHE has been reformulated in the language of geometric phase, \textit{i.e.} the Berry phase\cite{nagaosa2010anomalous}. Intrinsic contribution of the anomalous Hall conductivity (AHC) can be calculated through Kubo formula in its spectral representation, which is mathematically identical to the Berry phase representation\cite{yao2004first,nagaosa2010anomalous}. In analogy to AHC, intrinsic contribution of the spin Hall conductivity (SHC) can be also calculated using the Kubo formula by replacing the electron velocity operator with a spin current velocity operator\cite{Guo2008,Sun2005,Shi2006}. In this context, it is widely accepted that the SHE and AHE share the same theoretical framework. 


In this paper, we study the SHE of ferromagnets using first principle calculations with the generalized Kubo formula. We use bcc-Fe and fcc-Ni as prototypes of the ferromagnetic system. One of the key features in ferromagnets is the existence of large exchange interaction, which breaks the $SU(2)$ rotation symmetry in the spinor space. 
We release the constraint of parallel configuration between the spontaneous magnetization and the spin quantization axis of the conduction electrons\cite{nakamura2003enhancement}. 

We find the magnitude of the intrinsic spin Hall conductivity (SHC) in Fe and Ni can be varied via changes in the magnetization direction with respect to the spin polarization of the conduction electrons. 
The spin Berry curvature and the Berry curvature are mapped in the momentum space together with the spin character of the corresponding bands to study their correlation.


\begin{figure}[t]
\includegraphics[width=8cm]{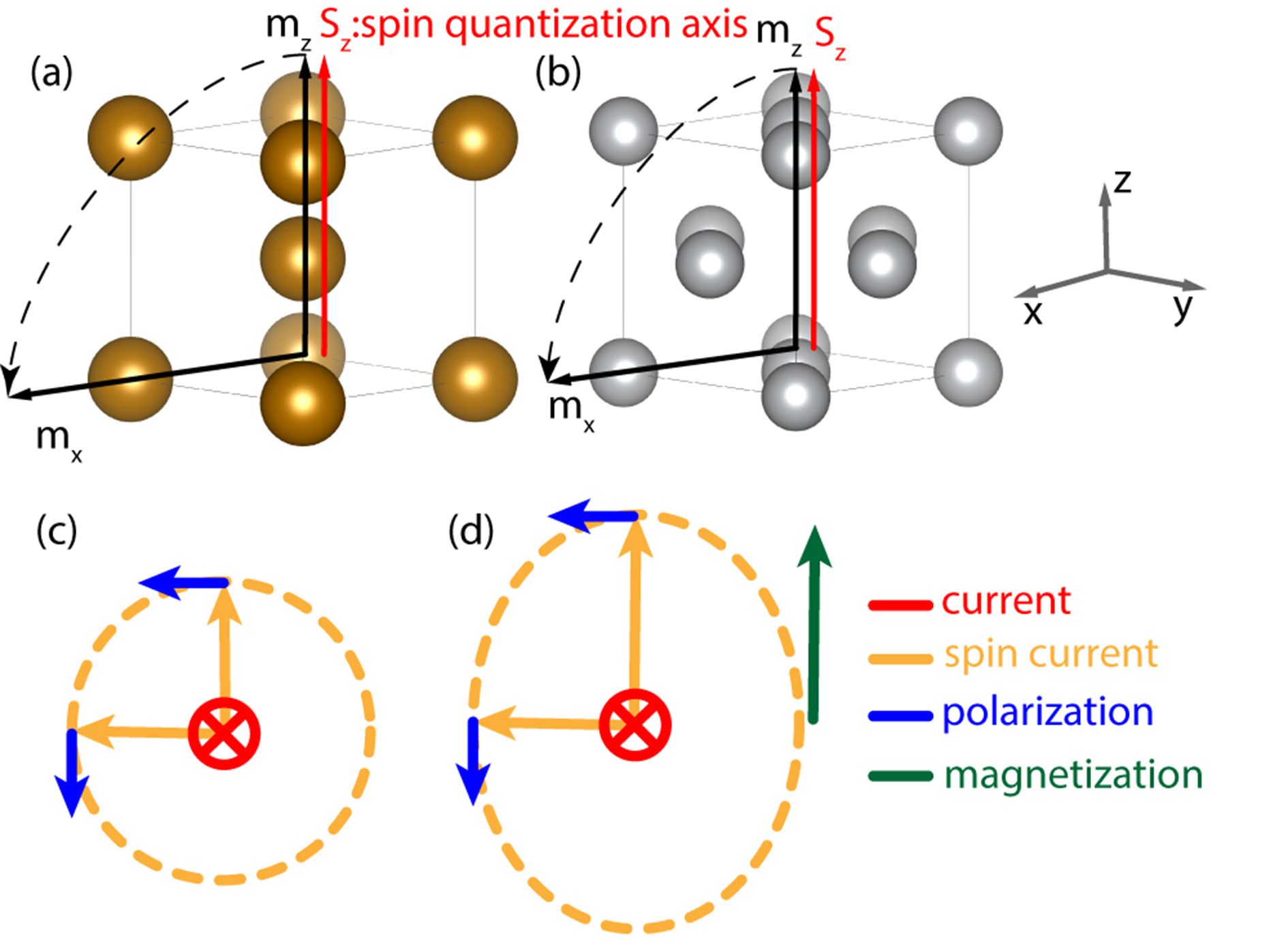}%
\caption{\label{fig:1} (a,b) Crystal structure of bcc-Fe (a) and fcc-Ni (b). Definition of the coordinate axis, with respect to the crystal structure, is shown on the right. (c,d) Schematic illustration of the magnetization direction dependent spin Hall effect. The red, blue and orange arrows indicate directions of current, polarization and flow of the spin current, respectively. The size of the orange arrows illustrates the magnitude of the spin Hall conductivity (SHC). In paramagnets (c), the magnitude of SHC is symmetric and does not depend on the electron spin direction. In ferromagnets (d), the SHC depends on the polarization of the electron spin with respect to the magnetization direction. The green arrow indicates the magnetization direction.}
\label{fig:1}
\end{figure}



\section{Calculation model}
The density functional theory (DFT) calculations is performed by full-potential linearized augmented-plane-wave method (FLAPW)\cite{wimmer1981wimmer,weinert1982total,nakamura2003enhancement} with generalized gradient approximation (GGA)\cite{perdew1996generalized} for the exchange correlation. The primitive cell of bcc-Fe and fcc-Ni are constructed with lattice constant chosen from experimentally determined values, $a_{Fe}=2.86 \text{\AA}, a_{Ni}=3.52 \text{\AA}$\cite{yelsukov1992mossbauer,van1994laser}. Muffin-tin (MT) radius are taken 2.2 bohrs for both Fe and Ni, respectively. The angular momentum expansion inside MT spheres is truncated at $l=8$ for the wave functions, charge and spin densities, and potential. The reciprocal ($k$-) space is divided into $16 \times 16 \times 16$ meshes for calculating charge and spin densities. The spin orbit coupling (SOC) is treated via the second variational method. 

The eigenstates are represented by a linear combination of the LAPW basis functions multiplied with a spinor\cite{nakamura2003enhancement}:
\begin{equation}
\Psi_{n,k}(\mathbf{r}) = \sum^{G_{\textrm{max}}}_{q=k+G} C_{n,q} \psi_{n,q} (\mathbf{r}) \chi_{n,q}. \label{eqn:1}
\end{equation}
$\psi_{n,q} (\mathbf{r})$ is the LAPW basis function, $\chi_{n,q}$ is the two component spinor that represents the spin direction of the state $(q,n)$, and $C_{n,q}$ is the expansion coefficient. $G$ is the reciprocal lattice vector. LAPW functions have a cutoff: $G_{\textrm{max}}$ is the cutoff vector with $|G_{\textrm{max}}| = 3.9 \text{ a.u.}^{-1} $. The electron density,
\begin{equation}
\rho_{\alpha}(\mathbf{r}) = \sum_{k} \sum_{n \in \textrm{occ}} \Psi^{\dagger}_{n,k}(\mathbf{r}) \sigma_{\alpha}\Psi_{n,k}(\mathbf{r}), %
\label{eqn:2} 
\end{equation}
contains a $U(1)$ part and a $SU(2)$ part, \textit{i.e.} $\rho_{\alpha} (\mathbf{r})=(\rho_{0} (\mathbf{r}), m_{k} (\mathbf{r}) )$, where $ \rho_{0} (\mathbf{r}) $ and $ m_{k} (\mathbf{r}) $ correspond to charge density and spin density, respectively. $ \sigma_{\alpha}$ represents the generalized Pauli matrix which has four components in our convention, $\sigma_{\alpha}=( I_2,\sigma_{k} )$. $I_2$ is a $2 \times 2$ unit matrix, $\sigma_{k}$ is the Pauli matrix, the greek indices (\textit{e.g.} $\alpha$, $\beta$) run from $0$ to $3$ representing four vectors, and the roman indices (\textit{e.g.} $i,j,k$) correspond to space coordinates (\textit{i.e.} $x,y,z$). The summation over $n$ in Eq.~(\ref{eqn:2}) imply summation over all occupied bands.

The Kohn-Sham single-particle Hamiltonian is written as
\begin{eqnarray}
\mathcal{H} =\mathcal{H}_{\textrm{kin}} I_2  +V_{\alpha} ( \mathbf{r})   \sigma_{\alpha}.
\label{eqn:3}
\end{eqnarray}
$\mathcal{H}_{\textrm{kin}}$ is kinetic energy term and $V_{\alpha}=( V_0(\mathbf{r}),V_k(\mathbf{r}) )$ is the effective potential with a non-magnetic $U(1)$ part and a magnetic $SU(2)$ part. In our calculation, the spin quantization axis is fixed along the $z$-axis in the Cartesian coordinate; that is, spin current with polarization along the $z$-axis is studied. Computations are carried out for two different magnetic configurations, noted as $m_z$ and $m_x$, by setting the magnetic moment along $z$ and $x$, respectively. (One may vary the spin quantization axis while keeping the magnetization direction fixed, which will return the same results.) Bcc-Fe and fcc-Ni are chosen as prototypes of ferromagnets which do not exhibit significant crystalline anisotropy.

The intrinsic anomalous Hall conductivity (AHC) and the intrinsic spin Hall conductivity (SHC) are obtained from the linear response Kubo formula in the static limit \cite{fang2003anomalous,yao2004first,PhysRevLett.92.126603,Guo2008,FENG2016428}. 
We define the off-diagonal conductivity tensor $\sigma_{ij}^{(\alpha)}$ as
\begin{eqnarray}
\sigma_{ij}^{(\alpha)} &=& -e^2 \hbar \int_{BZ} \frac{d^3 \mathbf{k} }{(2 \pi)^3} \Omega_{ij}^{(\alpha)}( \mathbf{k} ) \\
\label{eqn:4}
\Omega_{ij}^{(\alpha)}( \mathbf{k} ) &=& -\sum_{n' \neq n} \Big[ f(\epsilon_n( \mathbf{k} )) -f(\epsilon_n'( \mathbf{k} )) \Big] \notag \\
& & \times \frac{ \text{Im} \Big[ \bra{ n,\mathbf{k}} \hat{v}_i^{(\alpha)} \ket{ n',\mathbf{k}}  \bra{n',\mathbf{k}} \hat{v}_j^{(0)} \ket{ n,\mathbf{k}} \Big] }{\Big(\epsilon_{n}( \mathbf{k} ) - \epsilon_{n'}( \mathbf{k} ) \Big)^2} \label{eqn:5}
\end{eqnarray}
where $\sigma_{ij}^{(0)}$ and $\sigma_{ij}^{(k)}$ are the AHC and SHC, respectively, and $\Omega_{ij}^{(\alpha)}$ represents the generalized Berry curvature: $\Omega_{ij}^{(0)}$ and $\Omega_{ij}^{(k)}$ are the Berry and spin Berry curvatures. $v_i^{(\alpha)}= \frac{1}{2} \{ \sigma_{\alpha}, v_i\}$ is the general velocity operator with the subscript and superscript denoting the spatial coordinate and the spin quantization axis\cite{Jin2006}, respectively. $v_i^{(0)}$ and $v_i^{(k)}$ are the charge velocity and spin velocity operators. For the latter, we compute $v_i^{(3)}$ since the spin quantization axis is fixed along the $z$-axis in the calculations. With the zero-temperature assumption employed in our calculations, the Fermi distribution $f(\epsilon( \mathbf{k} ))$ reduces to a step function $\Theta(\epsilon_F-\epsilon( \mathbf{k} ))$ ($\epsilon_F$ is the Fermi energy).

To reduce numerical error, we extend the size of $k$-point mesh up to $64 \times 64 \times 64$ with total of 262,144 special $k$ points inside the first BZ to calculate the SHC. 
We use the internal coordinate system when expressing the wave vectors in the reciprocal space, \textit{i.e.} the $\bm{k}$ vectors are expressed in units of $\frac{\pi}{a}$, where $a$ is the lattice constant.


\section{Results and discussions}
The AHC and SHC for the two magnetization configurations are presented in Table.~\ref{tab:1}. The values of the intrinsic AHC of bcc-Fe and fcc-Ni show good agreement with previous reports\cite{yao2004first,fuh2011intrinsic}.
For both bcc-Fe and fcc-Ni, the non-vanishing component of the AHC changes from $\sigma_{yx}^{(0)}$ to $\sigma_{zy}^{(0)}$ when the magnetization direction is changed from being parallel to the $z$ axis to the $x$ axis. The magnitude of the non-vanishing component of the AHC remains the same. We have also studied the total magnetic moment of the system, which are found to be identical for both magnetization directions. 
These results are consistent with the symmetry of cubic systems. 

\begin{table}[b]
   \begin{ruledtabular}
   \caption{ Calculated anomalous Hall conductivity, spin Hall conductivity, and total spin magnetic moment ($\mu_{B}$) of bcc-Fe and fcc-Ni. The unit of AHC and SHC are S/cm and ($\frac{\hbar}{e}$) S/cm, respectively. }
   \label{tab:1}
   \begin{tabular}{cccccc}
  \textrm{Sample}&
 \textrm{$\sigma_{yx}^{(0)}$}&
  \textrm{$\sigma_{zy}^{(0)}$}&
 \textrm{$\sigma_{yx}^{(3)}$}&
 \textrm{$\sigma_{zy}^{(3)}$}&
  \textrm{$m_{tot}$}\\
\colrule
$\text{bcc-Fe ($m_z$)}$ & 747  & 0 & 130 & 0 &2.18\\
$\text{bcc-Fe ($m_x$)}$ & 0 & 747 & 527  & 0 &2.18\\
$\text{fcc-Ni ($m_z$)}$ & -2414 & 0 & 1535 &0 &0.60 \\
$\text{fcc-Ni ($m_x$)}$ & 0 & -2414 &  2358 &0 &0.60 \\
   \end{tabular}
  \end{ruledtabular}
\end{table}


\begin{figure}[h]
 \includegraphics[width=8cm]{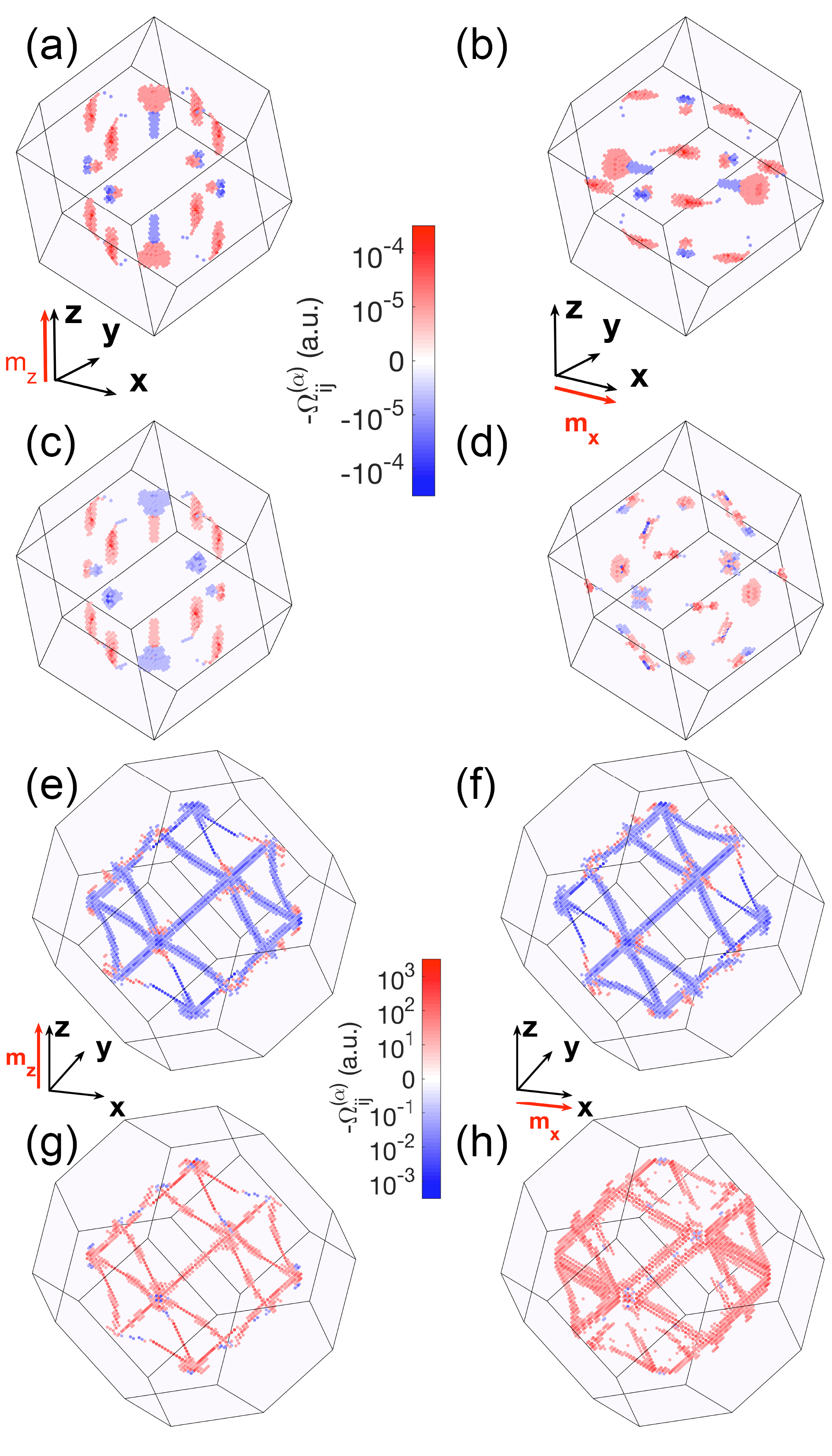}%
 \caption{\label{fig:2} (a-h) Berry and spin Berry curvatures projected on first Brillouin zone with two different magnetic configurations. The left and right panels present results when the magnetization points along $z$ and $x$, respectively. (a-d) show calculation results for bcc-Fe, (e-h) display those for fcc-Ni. (a,e) $\Omega_{yx}^{(0)}$, (b,f) $\Omega_{zy}^{(0)}$, (c,d,g,h) $\Omega_{yx}^{(3)}$.}
 \label{fig:2}
 \end{figure}

In contrast, the non-vanishing component of SHC, $\sigma_{yx}^{(3)}$, changes when the magnetization direction is changed. Note that here the polarization of the spin current is fixed along the $z$ axis\cite{seemann2015symmetry} regardless of the magnetization direction. (If we were to choose the spin quantization axis to follow the magnetization direction, as is done for many cases, the SHC will be invariant for cubic systems.) Recent studies have shown that such transverse spin current in ferromagnets, \textit{i.e.}, spin current with polarization orthogonal to the magnetization, does not dephase and persists in typical ferromagnets\cite{amin2019}. 
Interestingly, the magnitude of SHC considerably varies with changes in the magnetization direction. For example, in bcc-Fe, the SHC changes from 130 ($\frac{e}{\hbar}$) S/cm to 520 ($\frac{\hbar}{e}$) S/cm when the magnetization direction is rotated from $z$ to $x$. Such magnetization direction dependent SHC in ferromagnets is the main findings of this paper. 

To analyze the change in SHC with respect to the magnetization direction ($m_z$ and $m_x$), we first show stereoscopic mapping of the non-vanishing components of the Berry curvature ($\Omega_{yx}^{(0)}$) and the spin Berry curvature ($\Omega_{yx}^{(3)}$) projected in the first Brillouin zone. 
For bcc-Fe, upon rotating the magnetization from $z$ to $x$, the Berry curvature rotates following the the profile of the band structure [Figs.~\ref{fig:2}(a,b)]. The spin Berry curvature, in contrast, changes its profile upon rotation of magnetization [Figs.~\ref{fig:2}(c,d)].
The trend is the same for for fcc-Ni: $\Omega_{yx}^{(0)}$ rotates along with the band structure [Figs.~\ref{fig:2}(e,f)] when the magnetization direction is changed from $z$ to $x$ whereas $\Omega_{yx}^{(3)}$ changes its profile [Figs.~\ref{fig:2}(g,h)]. The region of non-zero $\Omega_{yx}^{(3)}$ notably increases when the magnetization is rotated, which causes the increase in the SHC.

\begin{figure}[h]
\includegraphics[width=9cm]{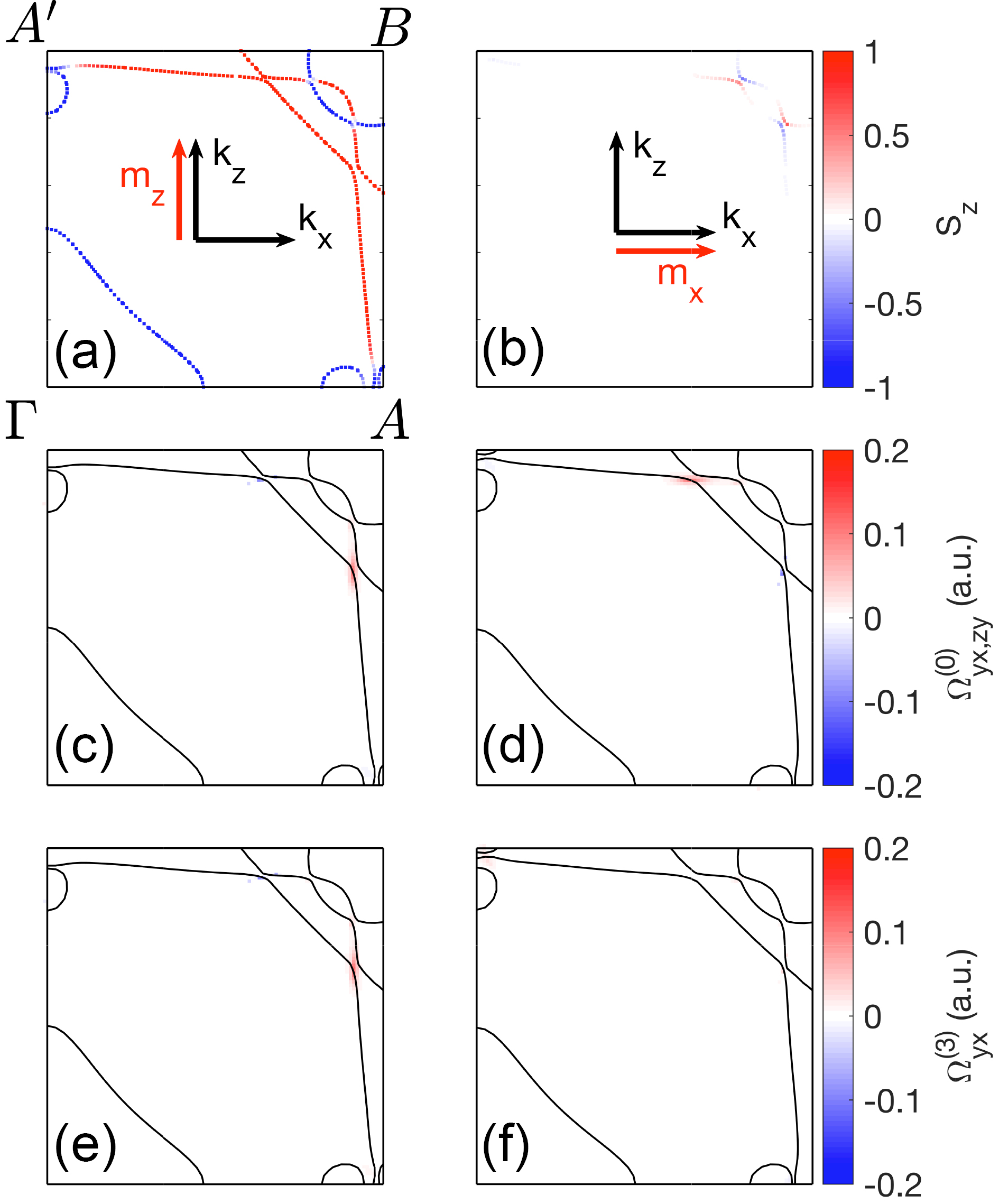}%
\caption{\label{fig:3} Reciprocal plane $\Gamma$(0,0,0)-$A$($\frac{1}{2}$,0,0)-$A'$(0,0,$\frac{1}{2}$)-$B$($\frac{1}{2}$,0,$\frac{1}{2}$) plotted with (a,b) spin expectation value $s_z,$. (c,d) Berry curvatures $\Omega_{yx}^{(0)}$ and $\Omega_{zy}^{(0)}$. (e,f) spin Berry curvature $\Omega_{yx}^{(3)}$. The magnetization points along $z$ (a,c,e,g) and $x$ (b,d,f,h). The Fermi contour of selected plane in Black lines (a-d) and red-blue lines (e-h) coded by spin character.  }
\label{fig:3}
\end{figure}


As in previous studies\cite{yao2004first,Guo2008}, the Berry curvature and the spin Berry curvature are enhanced at points in the reciprocal space where near degenerate states are lifted to form an avoided crossing, \textit{e.g} by spin orbit coupling. To articulate the change of SHC under rotation of the magnetization, we choose two prototypical avoided crossings, labeled as Type I and Type II. Type I (II) represents degenerate states with the polarization of the electron spin being parallel (anti-parallel). We find the change in $\Omega_{yx}^{(3)}$ with magnetization rotation is qualitatively different for the two types of avoided crossings. 

Characteristics of a representative avoided crossing of Type I is shown in Fig.~\ref{fig:3}. Here we show the projection of the spin character, the Berry curvature and the spin Berry curvature on the $k_x$ - $k_z$ plane ($\Gamma$(0,0,0)- $A$($\frac{1}{2}$,0,0) -$A'$(0,0,$\frac{1}{2}$) plane).
As the $z$-component of the spin is plotted in Figs.~\ref{fig:3}(a) and \ref{fig:3}(b), the color plot shows almost no contrast when the magnetization points along the $x$-axis (Fig.~\ref{fig:3}(b)) since the polarization for a large number of states is parallel to the magnetization.
We focus on the two avoided crossing points in the upper right corner of Fig.~\ref{fig:3}(a), where the states involved posses the same (majority) spin character.
The Berry curvature for magnetization along $z$ ($\Omega_{yx}^{(0)}$) and $x$ ($\Omega_{zy}^{(0)}$) is shown in Fig.~\ref{fig:3}(c) and \ref{fig:3}(d), respectively. 
$\Omega_{yx}^{(0)}$ and $\Omega_{zy}^{(0)}$ exhibit identical magnitude at the corresponding avoided crossing point, consistent with the symmetry of the AHE\cite{seemann2015symmetry}. 
(Note that here we are showing one quadrant of the $k_x$-$k_z$ plane of the Brillouin zone and the crystal structure has mirror symmetry alone $k_y$-$k_z$ plane.) 
The spin Berry curvature, in contrast, shows significant change in its magnitude upon rotation of the magnetization. 
Whereas a sizable contribution to $\Omega_{yx}^{(3)}$ is found at one of the avoided crossing point when the magnetization points along $z$ (Fig.~\ref{fig:3}(c)), the corresponding $\Omega_{yx}^{(3)}$ vanishes when the magnetization is rotated to $x$ (Fig.~\ref{fig:3}(d)).
We also note that the Berry and the spin Berry curvatures are nearly identical when the magnetization points along $z$, however, the relation does not hold when the magnetization is rotated to $x$.
Overall, in Type I crossing, we find the magnitude of $\Omega_{yx}^{(3)}$ changes from a finite value (positive or negative) to near zero upon rotating the magnetization from $z$ to $x$.

A representative avoided crossing of Type II is shown in Fig.~\ref{fig:4}.
Here we show projection of the corresponding properties on the same $k_x$ - $k_z$ plane as in Fig.~\ref{fig:3} but shifted along $k_y$ by $\frac{1}{2}$. 
In this plane, there are two avoided crossings at the top left and bottom right corners.
As evident in Fig.~\ref{fig:3}(a), the crossings are characterized by a pair of bands with opposite spin.
For Type II crossing, the spin character rapidly changes at the crossing point from majority to minority spins and vice versa.
Thus $s_z$ of the states involved at the crossing are nearly zero when the magnetization points along $z$, which is in contrast to the spin states involved in the Type I avoided crossing points.
For the Berry curvature, again we find it exhibits a two-fold symmetry: upon rotation of the magnetization direction from $z$ to $x$, $\Omega_{yx}^{(0)}$ (Fig.~\ref{fig:4}(c)) and $\Omega_{zy}^{(0)}$ (Fig.~\ref{fig:4}(d)) exhibit the same magnitude at the corresponding avoided crossing points.

The characteristics of the spin Berry curvature, however, is different compared to that of the Type I crossing.  
Interestingly, we find $\Omega_{yx}^{(3)}$ is nearly zero for magnetization along $z$ (Fig.~\ref{fig:4}(e)) whereas it shows a large positive value when the magnetization is directed along $x$ (Fig.~\ref{fig:4}(f)).
It is the large $\Omega_{yx}^{(3)}$ found at one of the avoided crossings (bottom right of Fig.~\ref{fig:4}(f) and the related points in the Brillouin zone) that contributes to the large SHC when the magnetization points along $x$. 
Note that correlation between $\Omega_{yx}^{(0)}$ and $\Omega_{yx}^{(3)}$ found in Type I crossing (magnetization along $z$) is lost in Type II crossing. 
In the following, we provide a simple analytical formula that relates the spin character of the states involved in the crossing and the size of Berry and spin Berry curvatures. 

\begin{figure}[h]
\includegraphics[width=9cm]{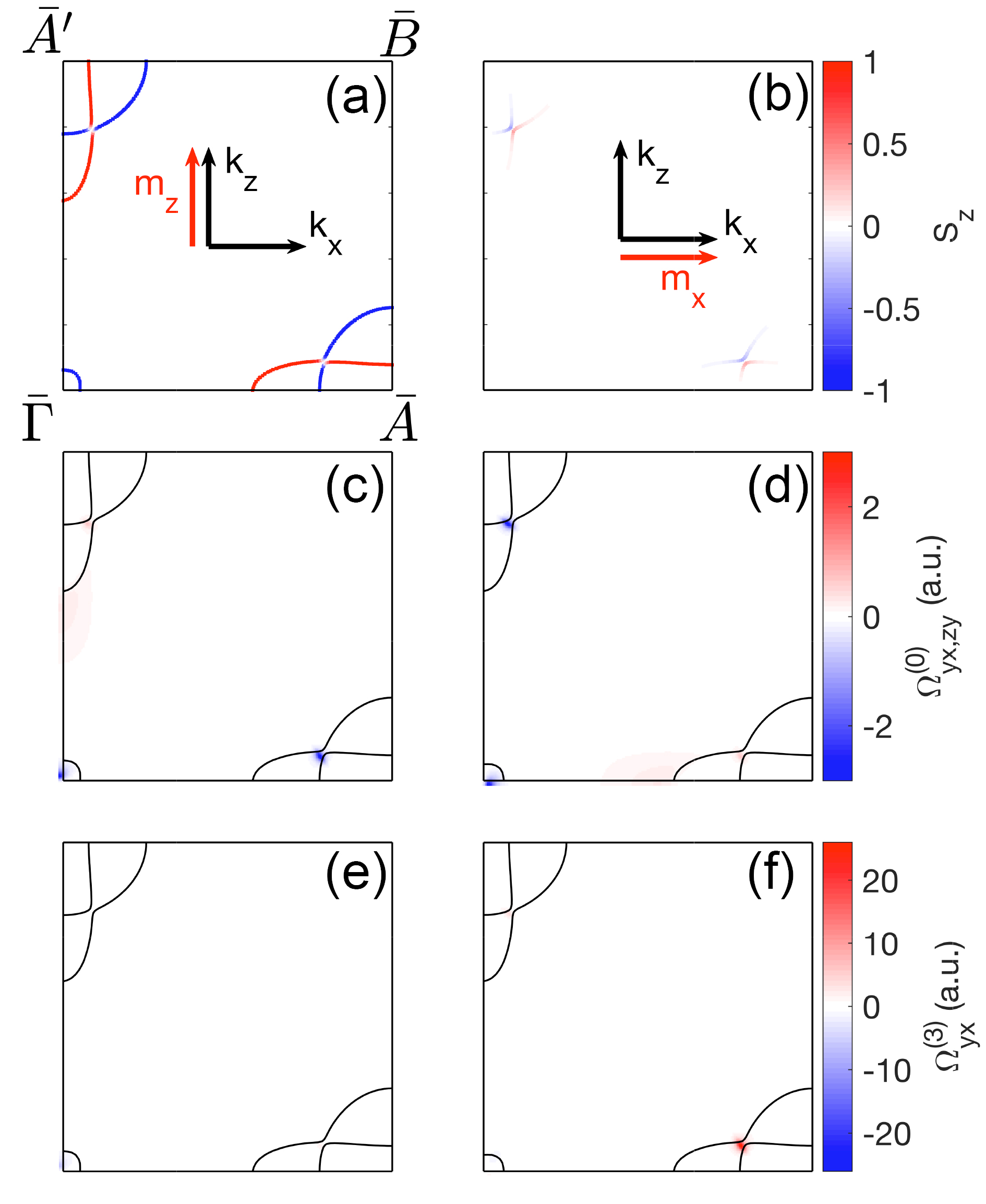}%
\caption{\label{fig:4} Reciprocal plane $\bar{\Gamma}$(0,$\frac{1}{2}$,0)-$\bar{A}$($\frac{1}{2}$,$\frac{1}{2}$,0)-$\bar{A}'$(0,$\frac{1}{2}$,$\frac{1}{2}$)-$\bar{B}$($\frac{1}{2}$,$\frac{1}{2}$,$\frac{1}{2}$) plotted with (a,b) spin expectation value $s_z,$. (c,d) Berry curvatures $\Omega_{yx}^{(0)}$ and $\Omega_{zy}^{(0)}$. (e,f) spin Berry curvature $\Omega_{yx}^{(3)}$. The magnetization points along $z$ (a,c,e,g) and $x$ (b,d,f,h). The Fermi contour of selected plane in Black lines (a-d) and red-blue lines (e-h) coded by spin character.  }
\label{fig:4}
\end{figure}

To describe the characteristics of Berry and spin Berry curvatures in ferromagnets, we model the system using the Bloch wave functions.
Let us assume a general spinor for the Bloch state when the magnetization points along $+z$, \textit{i.e.}  
$ \ket{ n,\mathbf{k}}_{z} = C_n (\mathbf{k}) \left[\begin{array}{c} \cos \frac{\theta_n (\mathbf{k})}{2} \\ \sin \frac{\theta_n (\mathbf{k})}{2} \end{array}\right] $, where $\theta_n (\mathbf{k})$ represents the polar angle of the electron spin with respect to the magnetization direction. 
Upon rotation of the magnetization from $+z$ to $+x$, the electron spin rotates its direction due to the $s$-$d$ exchange coupling. For magnetization pointing along $+x$, the Bloch state becomes: $ \ket{ n,\mathbf{k}}_{x} = C_n (\mathbf{k}) \left[\begin{array}{c} \cos ( \frac{\theta_n (\mathbf{k})}{2}+\frac{\pi}{4} ) \\ \sin ( \frac{\theta_n (\mathbf{k})}{2}+\frac{\pi}{4} ) \end{array}\right] $.
Note that $\theta_n (\mathbf{k})$ depends on the wave vector $\mathbf{k}$ and the band index $n$.
For example, when the majority and minority spin bands cross (\textit{i.e.}, Type II crossing), the electron spin direction at the crossing point is orthogonal to the magnetization due to the finite spin orbit coupling of the host material. 
This causes a non-zero $\Omega_{yx}^{(3)}$ when the magnetization points along $x$, as will be evident below.

The matrix elements of $\hat{v}_{i}^{(0)}$ and $\hat{v}_{i}^{(3)}$ can be calculated as
\begin{eqnarray}
\label{eq:vi0z}
\langle \hat{v}_{i}^{(0)} \rangle_{z} \equiv \bra{  n,\mathbf{k}}_{z} \hat{v}_{i}^{(0)} \ket{  n',\mathbf{k}}_{z} &=&  \langle v_i( \mathbf{k}) \rangle\cos \frac{\theta' -\theta}{2}\\
\label{eq:vi0x}
\langle \hat{v}_{i}^{(0)} \rangle_{x}  \equiv \bra{  n,\mathbf{k}}_{x} \hat{v}_{i}^{(0)} \ket{  n',\mathbf{k}}_{x} &=&  \langle v_i( \mathbf{k}) \rangle\cos \frac{\theta' -\theta}{2}\\
\label{eq:vi3z}
\langle \hat{v}_{i}^{(3)} \rangle_{z} \equiv \bra{  n,\mathbf{k}}_{z} \hat{v}_{i}^{(3)} \ket{  n',\mathbf{k}}_{z}  &=& \langle v_i( \mathbf{k}) \rangle  \cos \frac{\theta' + \theta}{2} \\
\label{eq:vi3x}
\langle \hat{v}_{i}^{(3)} \rangle_{x} \equiv \bra{  n,\mathbf{k}}_{x} \hat{v}_{i}^{(3)} \ket{ n',\mathbf{k}}_{x}  &=&  \langle v_i( \mathbf{k}) \rangle \sin \frac{\theta' + \theta}{2} 
\end{eqnarray}
where $\theta$ and $\theta'$ are the average polar angle of the electron spin (averaged over all $\bm{k}$ states within a band) with respect to the magnetization direction for the two Bloch states $\ket{ n,\mathbf{k}}_{z}$ and $\ket{ n',\mathbf{k}}_{z}$, respectively. 
$\langle v_i( \mathbf{k}) \rangle$ represents the expectation value of the velocity operator with the spatial part of the Bloch state.
According to the Kubo formula (Eq.~(\ref{eqn:4})), the Berry curvature is represented by the product of two $U(1)$ velocity matrix elements, $\langle \hat{v}_{i}^{(0)} \rangle_{z,x} \langle \hat{v}_{j}^{(0)} \rangle_{z,x}$, whereas the spin Berry curvature is the product of $U(1)$ and $SU(2)$ velocity matrix elements, $\langle \hat{v}_{i}^{(3)} \rangle_{z,x} \langle \hat{v}_{j}^{(0)} \rangle_{z,x}$. 

Equations~(\ref{eq:vi0z}) - (\ref{eq:vi3x}) can account for the features found in Figs.~\ref{fig:3} and \ref{fig:4}.
First, for Type I crossing with majority spin states ($\theta = \theta' = 0$), \textit{i.e.} when the magnetization points along $z$, Eqs.~(\ref{eq:vi0z}) and (\ref{eq:vi3z}) indicate that the Berry curvature and the spin Berry curvature takes a positive value with the same magnitude.
When the magnetization is rotated to $x$, Eqs.~(\ref{eq:vi0x}) and (\ref{eq:vi3x}) suggest that the Berry curvature maintains its magnitude but the spin Berry curvature vanishes.
Such characteristics are in agreement with the calculation results shown in Fig.~\ref{fig:3}.
Equations~(\ref{eq:vi0z}) and (\ref{eq:vi3z}) also dictate that the signs of Berry and spin Berry curvature are opposite when the crossing is formed from minority spin states ($\theta = \theta' = \pi$).
The effect has been confirmed in the calculations.

For Type II crossing with mixed spin states ($\theta + \theta' \sim \pi$ and $\theta = \theta' \sim \frac{\pi}{2}$ at the crossing points ), Eqs.~(\ref{eq:vi0z}) and (\ref{eq:vi0x}) suggest that the Berry curvature is unchanged under magnetization rotation.
In contrast, the spin Berry curvature is expected to be zero when the magnetization points along $z$ and is signficantly enhanced when the magnetization is directed along $x$.
These suggestions are consistent with the results shown in Fig.~\ref{fig:4}.

The argument above shows that the spin Hall effect in ferromagnetic metals is dependent on the spin character of each band, \textit{i.e.} the spinor part of the Bloch function characterized by $\theta$. Symmetry of Berry curvature do not depends on the absolute spin character of states, $\theta$ or $\theta'$, but only on the relative angle of polarization between two states, $\theta'-\theta$. Since our analysis is based on the Kubo formula in which the Berry and spin Berry curvatures are described using two bands, we cannot assess the symmetry of Berry and spin Berry curvatures one particular band using its spin character. Further investigation is required to develop a model to make such prediction.

It is convenient to discuss AHE and SHE in the same framework of a $U(1) \times SU(2)$ theory, as purposed previously\cite{Zhang2006,Jin2006,Leurs2008,qu2019symmetry}. 
For the AHE in ferromagnets, the system has almost always been treated with the spin quantization axis (\textit{i.e.}, the polarization of the electrons) aligned along the magnetization direction. 
Therefore, the $U(1) \times SU(2)$ theory reduces to a parallel $U(1)$ transport model\cite{Leurs2008}. 
For 3\textit{d} transition metals with such parallel configuration, a strong correlation between the Berry curvature and the spin Berry curvature is found, for which one may consider the $U(1) \times U(1)$ theory is a good approximation.
This is also possible because the 3\textit{d} transition metals with large exchange splitting do not possess large SOC that will mix the spinor states. 
However, for a transverse spin current in ferromagnets, \textit{i.e.}, when the spin quantization axis is rotated away from the magnetization direction, the $U(1) \times U(1)$ approximation is no longer valid to discuss the SHE. 

In conclusion, we have used bcc-Fe and fcc-Ni as a prototype systems to study AHC and SHC in ferromagnets. 
Whereas the magnitude of the non-vanishing component of the AHC in ferromagnets is independent on the magnetization direction, the non-vanishing component of the SHC is highly dependent on the relative angle between the magnetization and the conduction electron spin orientation. With the conduction electron spin orientation fixed along $z$, the SHC of bcc-Fe (fcc-Ni) increases by a factor of 4 (1.5) when the magnetization direction is rotated from $z$ to $x$. 
Such a magnetization direction dependent SHC originates from the anisotropy of the spin current operator in the spinor space: as the spinor part of the Bloch states changes upon rotating the magnetization direction away from the conduction electron spin orientation, the matrix elements of the spin current operator with the Bloch states vary.
We have developed a simple analytical form to characterize the relation between the spinor state of the electrons and the Berry and spin Berry curvatures.
The form can account for many of the features found in the calculations. 
These results show that the SHC in ferromagnets have an extra handle, \textit{i.e.} the magnetization direction, to control its magnitude. Further investigation is required to clarify the effect for the extrinsic contributions to the SHC.

\begin{acknowledgments}
\end{acknowledgments}

\bibliography{Spinrotation_100320}

\end{document}